\documentclass[prl,twocolumn,showpacs,superscriptaddress]{revtex4-1}  % default: %\bibliographystyle{apsrev4.bst}
\usepackage{natbib}
\usepackage{graphicx}
\usepackage{subfigure}
\usepackage{epsfig} % epsf no longer supported in 4.1
\usepackage{amsmath}

\usepackage{color}
\begin{document}

\title{Anderson mobility gap probed by dynamic coherent backscattering}
\author{L. A. Cobus} %\email{umcobus@cc.umanitoba.ca}
\affiliation{Department of Physics and Astronomy, University of Manitoba, Winnipeg, Manitoba R3T 2N2, Canada}
\author{S. E. Skipetrov}
\affiliation{Universit\'{e} Grenoble Alpes, LPMMC, F-38000 Grenoble, France}
\affiliation{CNRS, LPMMC, F-38000 Grenoble, France}
\author{A. Aubry}
\affiliation{Institut Langevin, ESPCI ParisTech, CNRS UMR 7587, Universit\'{e} Denis Diderot - Paris 7, 1 rue Jussieu, 75005 Paris, France}
\author{B. A. van Tiggelen}
\affiliation{Universit\'{e} Grenoble Alpes, LPMMC, F-38000 Grenoble, France}
\affiliation{CNRS, LPMMC, F-38000 Grenoble, France}
\author{A. Derode}
\affiliation{Institut Langevin, ESPCI ParisTech, CNRS UMR 7587, Universit\'{e} Denis Diderot - Paris 7, 1 rue Jussieu, 75005 Paris, France}
\author{J. H. Page}
\affiliation{Department of Physics and Astronomy, University of Manitoba, Winnipeg, Manitoba R3T 2N2, Canada}

\date{\today}

\begin{abstract}
We use dynamic coherent backscattering to study one of the Anderson mobility gaps in the vibrational spectrum of strongly disordered three-dimensional mesoglasses. Comparison of experimental results with the self-consistent theory of localization allows us to estimate the localization (correlation) length as a function of frequency in a wide spectral range covering bands of diffuse transport and a mobility gap delimited by two mobility edges. The results are corroborated by transmission measurements on one of our samples.
\end{abstract}

%\pacs{42.25.Dd, 43.20.Gp, 71.23.An}

\maketitle

A quantum particle is trapped in a three-dimensional (3D) disordered potential if its energy $E$ is lower than the so-called mobility edge (ME) $E_c$. As was discovered by Philip Anderson in 1958, quantum interferences may increase $E_c$ to values that are much larger than the classical percolation threshold, an energy below which a classical particle would be trapped \cite{Anderson1958,Abrahams2010}. The link between $E_c$ and the statistical properties of disorder has been recently studied in experiments with ultracold atoms in random optical potentials \cite{Kondov2011,Jendr2012NP}. In contrast to quantum particles, classical waves---light or sound---may be Anderson localized by disorder only in a band of intermediate energies (or frequencies), the impact of disorder becoming weak in both high- and low-frequency limits \cite{John1984,John1991}. One thus expects a mobility 'gap' delimited by two MEs instead of a single ME. This is due to the difference between dispersion relations of quantum and classical waves \cite{VanTiggelen1991,VanTiggelen1994}. Resonant scattering may further complicate the spectrum by shifting the mobility gap or splitting it into several narrower ones. Mobility gaps can also exist for quantum particles when the disordered potential is superimposed on a periodic one---a common situation for electrons in crystals with impurities \cite{Chaikin1995}. In the present Letter we report the first experimental observation of a mobility gap for classical waves. To this end we take full advantage of experimental techniques available for classical waves but very difficult, if not impossible, to put in practice for quantum particles and, in particular, for electrons in disordered conductors. We perform frequency-, time-, position- and angular-resolved ultrasonic reflection and transmission experiments in strongly disordered `mesoglasses'---elastic networks of brazed aluminum beads. The results are compared with the self-consistent theory of localization to precisely locate the two MEs and to estimate the localization length $\xi$ throughout the mobility gap. $\xi$ diverges at the MEs, as expected.

Among the many definitions of Anderson localization, two of them rely either on the exponential decay of eigenmodes at large distances or the vanishing of diffusion \cite{VanTiggelen1999}. However, strictly speaking, both only apply in an infinite disordered medium, and not in experiments which involve finite samples with often open boundaries. In the latter case, waves can leak through the sample boundaries to the surrounding medium; hence, the eigenmodes no longer decay exponentially at large distances (because waves propagate freely outside the sample), and the transport is no longer blocked completely, even though wave diffusion is suppressed exponentially. This is why important efforts were devoted in recent years to study signatures of Anderson localization in finite 3D samples that can be seen as representative portions of infinite disordered media in which waves would be Anderson localized. The most impressive successes were achieved for quantities measured in transmission where time- and position-resolved measurements of wave intensity allowed unambiguous observation of Anderson localization of elastic waves \cite{Hu2008}, without complications due to absorption. However, an important shortcoming of such measurements is the weakness of transmitted signals that decay exponentially with sample thickness $L$ making the regime of very strong localization $L/\xi \gg 1$ inaccessible. Even in the diffuse regime, the transmitted intensity may become so weak that the measured signal is dominated by other, presumably weak phenomena (e.g., nonlinear effects or fluorescence in optics) which can be misinterpreted as a signature of Anderson localization \cite{Sperling2015,Perspectives2016}.

To circumvent the difficulties of transmission experiments, we develop a new approach to Anderson localization of waves based on time- and angular-resolved reflection measurements. The total reflection coefficient of a thick disordered sample is close to unity because almost all the incident energy is reflected, allowing for comfortable signal levels even deep in the localized regime. For a plane wave incident upon a slab of weakly disordered medium, $k \ell \gg 1$, the average reflection coefficient $R(\theta)$ is known to be almost Lambertian, but with a two-fold enhancement within a narrow angular range $\Delta \theta \sim (k_0 \ell^*)^{-1}$ around the exact backscattering direction $\theta = 0$ \cite{Kuga1984,VanAlbada1985,Wolf1985,Akkermans1986,Akkermans2007}. Here $k$ and $k_0$ are the wave numbers inside and outside the sample, respectively, and $\ell$ and $\ell^*$ are the scattering and transport mean free paths. If the incident wave is a short pulse, the shape $R(\theta, t)$ of this coherent backscattering (CBS) peak evolves in time whereas its relative amplitude remains constant \cite{Vreeker1988,Bayer1993,Tourin1997}. The width $\Delta \theta$ of the CBS peak decreases with time according to $\Delta \theta^2 \propto 1/Dt$, where $D$ is the wave diffusion coefficient, as can be easily found from the solution of the diffusion equation \cite{Akkermans2007}. CBS is a very general phenomenon due to constructive interferences of partial waves that follow time-reversed paths in a disordered medium. It was observed for light in suspensions of small dielectric particles \cite{Kuga1984,VanAlbada1985,Wolf1985} and clouds of cold atoms \cite{Labeyrie1999}, sound \cite{Bayer1993,Tourin1997}, seismic \cite{Larose2004} and matter \cite{Jendr2012} waves. Being an interference phenomenon, CBS seems natural to use as a probe of Anderson localization. However, the stationary (time-integrated) CBS peak was predicted to be only weakly affected by localization effects, with the most pronounced effect being the rounding of its tip which can also be due to absorption \cite{VanTiggelen2000}. Optical experiments confirmed the rounding of the tip \cite{wiersma1997,schuurmans1999}, but the conclusion that this behaviour was caused by  Anderson localization of light \cite{schuurmans1999} was not supported by transmission measurements performed on the same or similar samples \cite{VanDerBeek2012,GomezRivas2001}. In this context, the dynamic CBS is more promising as a probe of Anderson localization because its shape is independent of absorption provided the absorption coefficient is spatially uniform on average, and its width $\Delta \theta$ explicitly depends on the diffusion coefficient $D$. In a different context, recent theoretical work suggests that dynamic CBS of cold atoms in a random potential may serve as a probe of Anderson transition \cite{Ghosh2015}.

\begin{figure}[t]
\begin{subfigure}{}
\centering
\includegraphics[width=4.2cm]{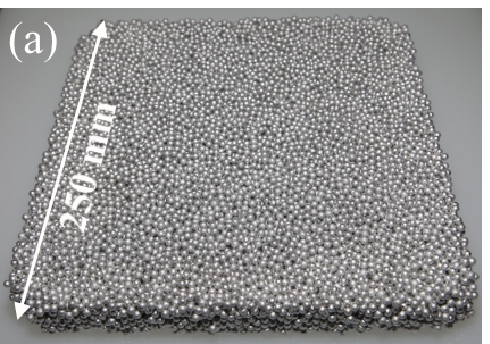}
\end{subfigure}
\begin{subfigure}{}
\centering
\includegraphics[width=3.8cm]{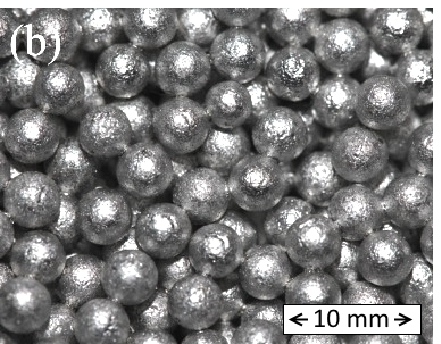}
\end{subfigure}
\begin{subfigure}
\centering
\includegraphics[width=8cm]{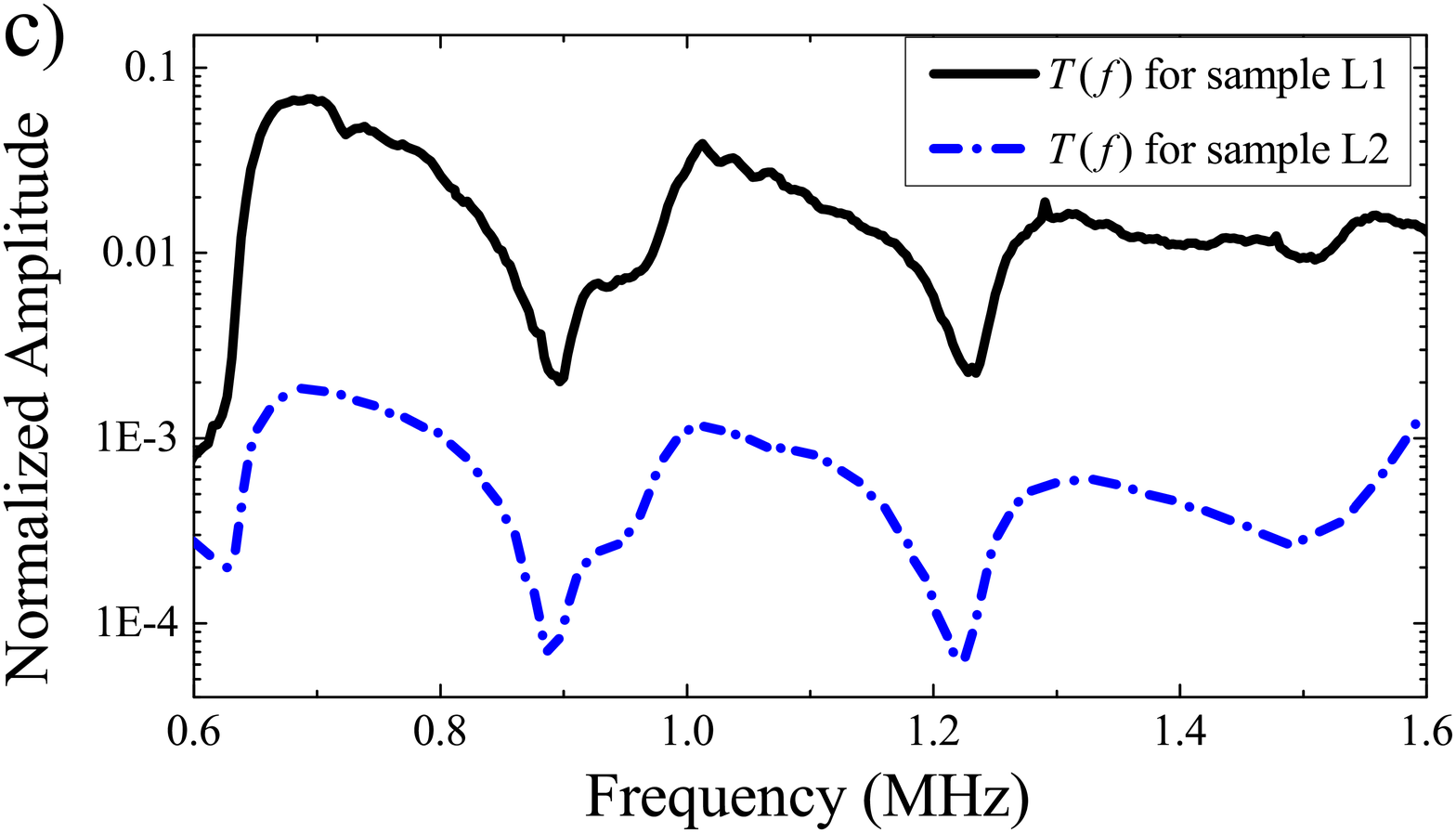}
\caption{(a)  Sample L1.  (b)  Bead structure of sample L1.  (c)  Amplitude transmission coefficient of ultrasonic waves through samples L1 and L2 as a function of frequency.}
\label{fig1}
\end{subfigure}
\end{figure}

In this Letter we report measurements of CBS from two of our mesoglass samples composed of aluminum beads brazed together (volume fraction $\sim 55$\%) to form an elastic network. The samples have the shape of slabs with cross-sections of 230$\times$250 mm$^2$ much larger than thicknesses $L_1= 25 \pm 2$ mm and $L_2= 38 \pm 2$ mm of samples L1 [see Fig.\ \ref{fig1}(a)] and L2, respectively.
They were waterproofed so that experiments could be performed in a water tank with immersion transducers or transducer arrays and the pores between the beads held under vacuum during all measurements. The samples are similar to those used in previous studies \cite{Hu2008,Hildebrand2014}, but instead of being monodisperse have a mean bead diameter of 3.93 mm with a polydispersity of about 20\%, which helps to randomize bead positions. The samples also have stronger elastic bonds between beads than previous samples, visible in Fig.\ \ref{fig1}(b). These differences influence the frequency dependence of amplitude transmission coefficient, shown in Fig.\ \ref{fig1}(c). Coupling between the individual resonances of the beads leads to frequency bands of relatively high transmission whose widths depend on the coupling strength \cite{Turner1998,Hu2008}, but these bands are narrow enough in our samples to cause transmission dips to appear in between.  The depth and width of the dips are lessened by the polydispersity and greater inter-bead bond strength compared with the monodisperse samples.

These dips may correspond to Anderson mobility gaps but one has to study the nature of wave transport in the corresponding frequency ranges to claim anything with certainty. Here we report a detailed study of wave transport around the transmission dip at 1.23 MHz. Ultrasound is very strongly scattered near this frequency; we have measured the product $k\ell$ as small as $k \ell \lesssim 3$. More details of sample L1 can be found in a previous work \cite{Aubry2014}. Sample L2 is too thick and too strongly scattering for many of the conventional methods of sample characterization in transmission to work. As no detectable coherent signal could be transmitted through L2 in the frequency range of interest, measurements of $k$ and $\ell$ from the coherent pulse \cite{Page1996} are not possible. However, both samples were fabricated using the same technique and have very similar composition, so that estimates from coherent measurements on sample L1 are expected to be a good approximation for L2 as well.

\begin{figure}[t]
\centering
\includegraphics[width=1.08\columnwidth]{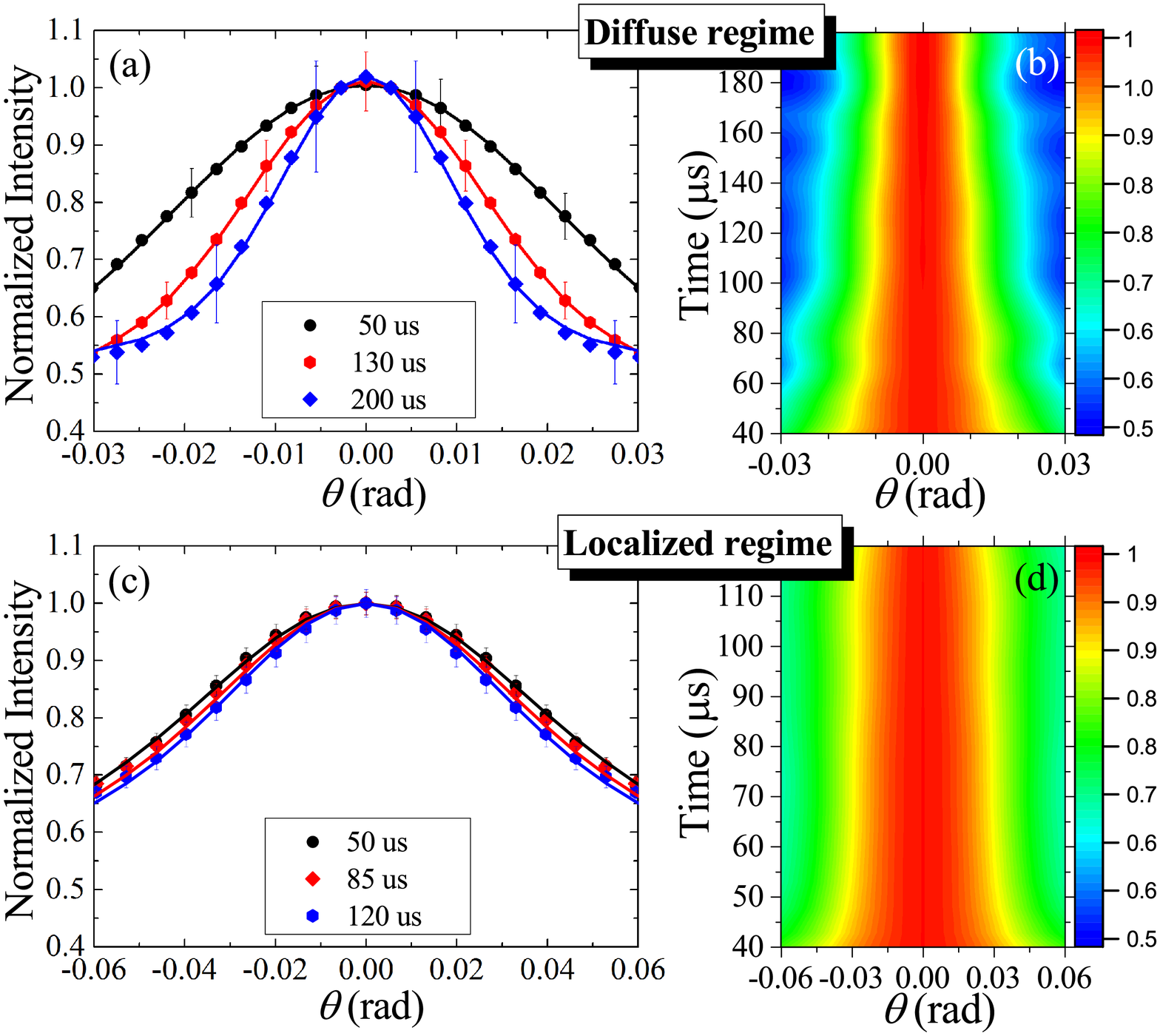}
\caption{Dynamic CBS profiles in the diffuse regime (1.65 MHz)  (a,b) and in the localized regime (1.22 MHz)  (c,d). The results in (a,b) are for sample L1, and in (c,d) for sample L2 (note the different angular scales).  In (a,c) theoretical fits (lines) and experimental data (symbols) are shown for three representative times. In (a), the data are fitted using diffusion theory, giving diffusion coefficient $D = D_B = 0.7$ mm$^2/\mu$s \cite{DiffTheoryNote}, whereas in (c) SC theory is used, giving $\xi = 16.5$ mm. Additional examples are shown in the Supplemental Material \cite{suppl}.  In (b,d) experimental CBS profiles are shown as a function of both time and angle. The profile narrows quite rapidly in the diffuse regime (b), but is almost constant over the accessible range of times in the localized regime (d). }
\label{fig2a}
\end{figure}

We measure the backscattered intensity using ultrasonic transducer arrays, placed in the diffuse far field of the samples (for details, see the Supplemental Material \cite{suppl}). A time-dependent `response matrix' was gathered by emitting with each element in turn, and recording the time-dependent backscattered field with all elements \cite{Aubry2014,Tourin1997}.  An average over configurations of disorder was performed by translating the array parallel to the sample surface and acquiring the response matrices for different positions. To obtain results as a function of both time $t$ and frequency $f$, the data were filtered using a Gaussian envelope of standard deviation 0.015 MHz, centered around $f$.
As has been previously reported \cite{Aubry2014}, these backscattering data show significant contributions from recurrent scattering due to the signal entering and leaving the sample near the same spot  \cite{Aubry2014,Wiersma1995}. Recurrent scattering complicates the analysis of CBS peaks, as it is difficult to determine the (roughly flat) background intensity level corresponding to large angles $\theta$. The recurrent scattering contribution was removed from the total backscattered intensity following the approach developed previously \cite{Aubry2014}.

To eliminate the effect of absorption, the time-dependent CBS profiles $R(\theta, t)$, where $\theta$ is the angle between source and receiver elements of the ultrasonic array, are normalized by $R(0,t)$ \cite{suppl}. Analogously to transverse confinement measurements in transmission \cite{Hu2008}, absorption cancels in the ratio $R(\theta,t)/R(0,t)$. Representative profiles $R(\theta,t)/R(0,t)$ are shown in Fig.~\ref{fig2a}.

To obtain a quantitative description of our data, we use the self-consistent (SC) theory of Anderson localization with a position-dependent diffusion coefficient $D(z, \Omega)$ presented in Refs.\ \cite{Skipetrov2006,Cherroret2008}. First, $D(z, \Omega)$ is determined from an iterative solution of the self-consistent equations for each depth $z$ inside the sample ($0 \leq z \leq L$). Second, the two-dimensional spatial Fourier transform of the intensity Green's function $C(q_{\perp}, z, z' = \ell_B^*, \Omega)$ is calculated using this $D(z, \Omega)$. Here $\ell_B^*$ is the transport mean free path in the absence of Anderson localization effects. Finally, the CBS profile $R(\theta, t)$ is obtained as a Fourier transform of $R(q_{\perp}, \Omega) = D(z=0, \Omega) \left. \partial C(q_{\perp}, z, z' = \ell_B^*, \Omega)/\partial z \right|_{z=0}$ where $q_{\perp} = k_0 \sin\theta$ \cite{suppl}. Fits to the experimental data obtained from this theory are shown in Fig.~\ref{fig2a}(c). We refer the reader to Ref.\ \cite{suppl} for the details of the fitting procedure. For a given frequency $f$, important outcomes of the fitting procedure are the location of $f$ with respect to the ME $f_c$ (indicating whether wave transport at $f$ is extended or localized) and the value of the localization length $\xi$ that characterizes the closeness to a ME and the extent of localization effects \cite{XiNote}.\

\begin{figure}[t]
\centering
\includegraphics[width=\columnwidth]{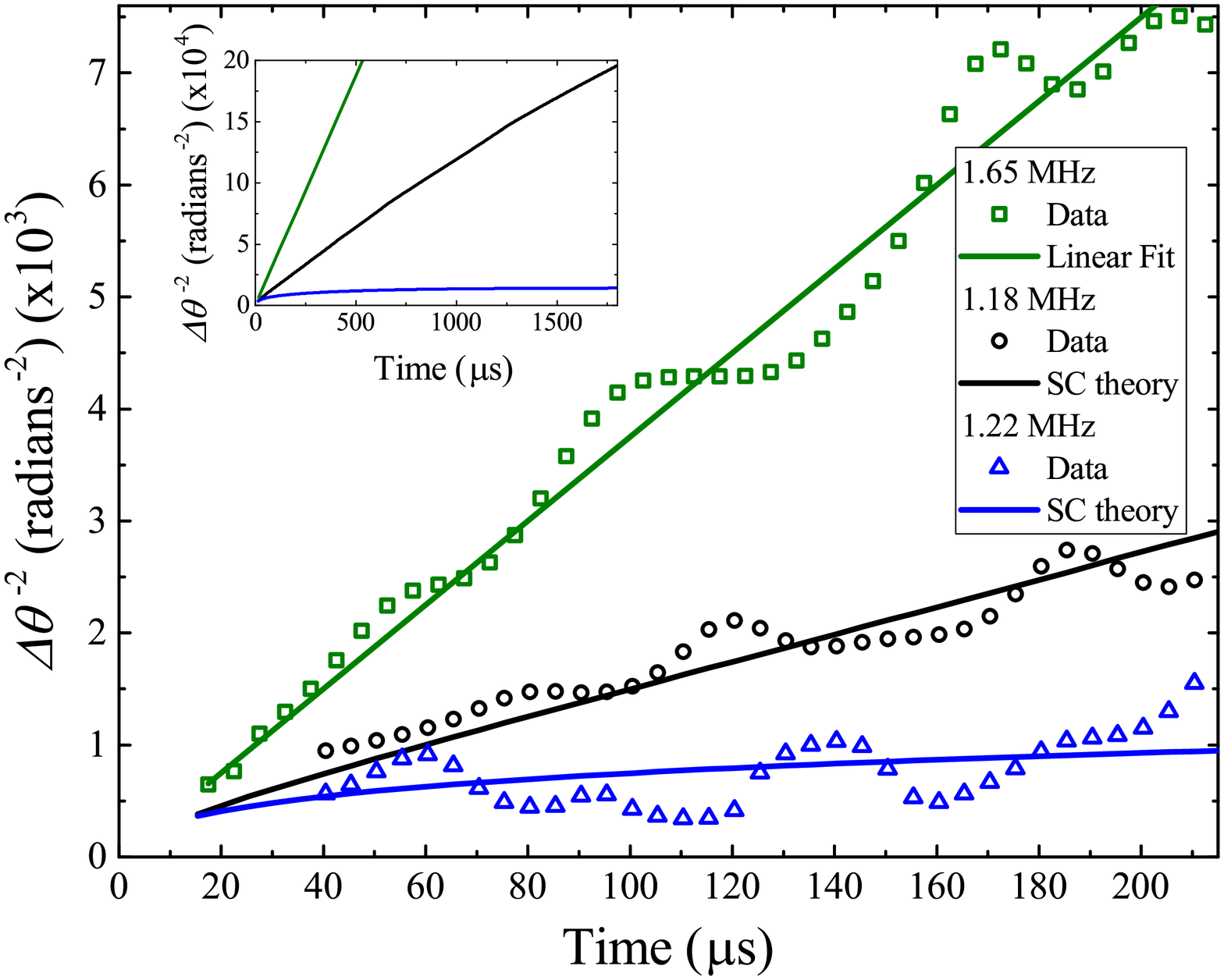}
\caption{Experimental results (symbols) and theoretical predictions (lines) for sample L1. Plotted is the reciprocal of the square of the half width at half maximum of the CBS peaks, $\Delta \theta^{-2}(t)$ (error bars are smaller than symbol sizes). Three representative frequencies are shown: $f = 1.65$ MHz (diffuse regime, diffusion coefficient $D_B = 0.7$ mm$^2/\mu$s extracted from the fit), $f = 1.18$ MHz (slower diffusion as a ME is approached; correlation length $\xi = 2.1$ mm), and $f = 1.22$ MHz (Anderson localization; localization length $\xi = 12.5$ mm). The inset shows theoretical predictions for longer times.}
\label{fig3}
\end{figure}

CBS profiles shown in Fig.\ \ref{fig2a}(a,b) exhibit the narrowing with time predicted by the diffusion theory. However, when approaching $f = 1.20$ MHz and beyond, the narrowing of CBS profiles slows down considerably (see Fig.~\ref{fig2a}(c,d) and Fig.\ S1 of Ref.\ \cite{suppl}). Such a slowing down is expected when a ME of the Anderson transition is approached and crossed because the width of the CBS peak $\Delta \theta$ behaves, roughly speaking, as the inverse width of the diffuse halo at the surface of the sample. The latter grows without limit in the diffuse regime but cannot exceed a value on the order of the localization length $\xi$ in the localized regime. Hence, the corresponding CBS profile stops shrinking and its width $\Delta \theta$ saturates. This is illustrated in Fig.\ \ref{fig3} where the different types of behavior can be clearly distinguished.

\begin{figure}[t]
\centering
\hspace*{-4mm}
\includegraphics[width=1.0\columnwidth]{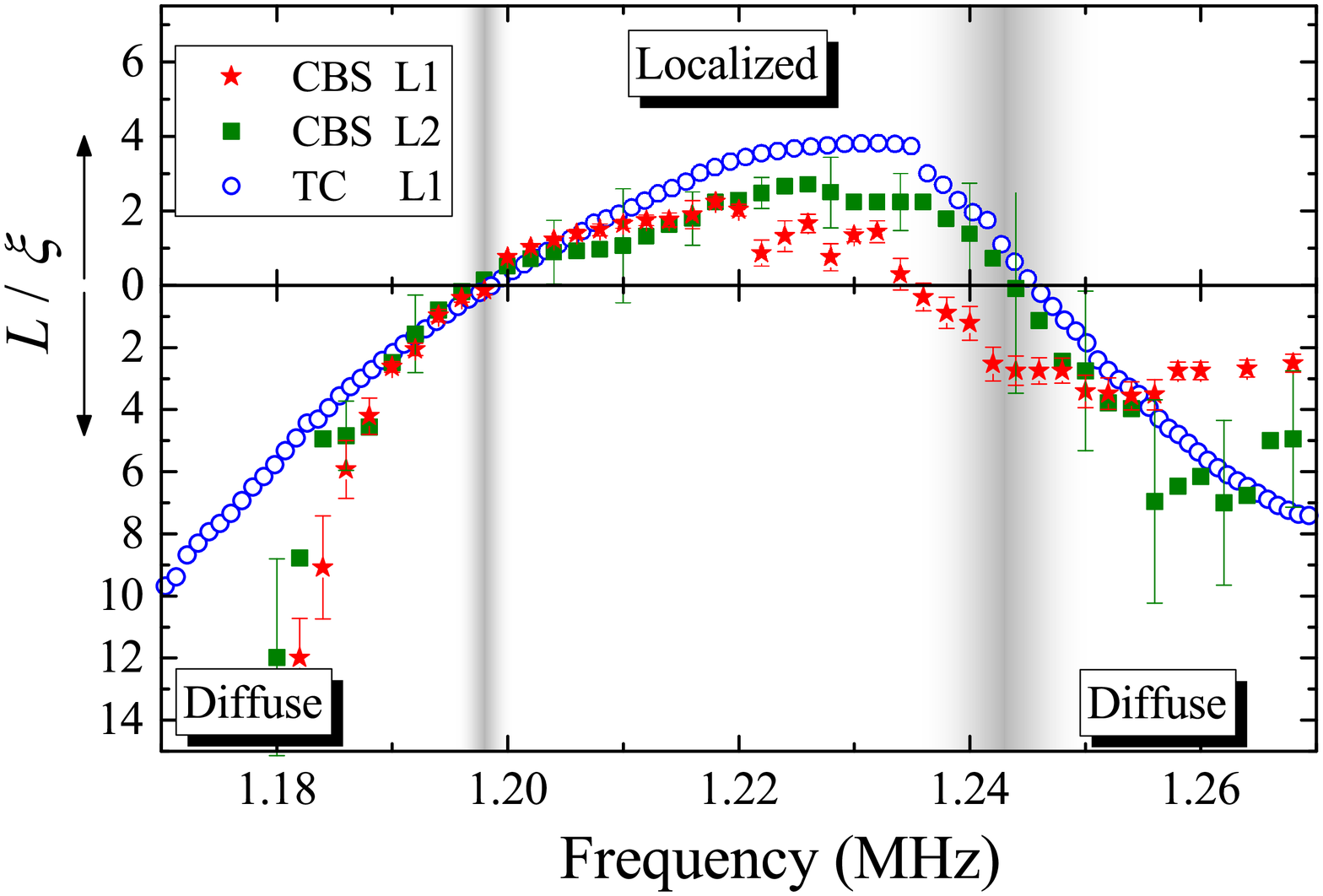}
\caption{The ratio of sample thickness $L$ to the localization (correlation) length $\xi$ obtained from fits to experimental CBS profiles (sample L1---red stars, sample L2---green squares) and transverse confinement (TC) data (sample L1---open circles). Error bars represent variations of $L/\xi$ that increase the reduced $\chi^2$ by unity; error bars are smaller than symbol size for transmission results. Fuzzy vertical gray lines show our estimates of mobility edges.}
\label{fig4}
\end{figure}

We performed systematic fits of SC theory to our data for frequencies from 1.17 to 1.27 MHz for both samples L1 and L2, thereby determining the frequency dependencies of the localization (correlation) length $\xi$. The results are shown in Fig.\ \ref{fig4} where MEs at approximately 1.20 and 1.24 MHz are indicated by fuzzy vertical gray lines. The Anderson mobility gap is clearly visible in between, whereas the wave transport is diffusive for frequencies below 1.20 and above 1.24 MHz. Other and possibly multiple mobility gaps can exist in our samples outside the frequency range from 1.17 to 1.27 MHz that we explored. It is important to note that although the position of the Anderson mobility gap that we have found coincides with one of the dips in the transmission spectra of Fig.\ \ref{fig1}(c), the latter is not sufficient to claim the existence of the former. Indeed, a dip in transmission can simply correspond to spectral regions with a low density of states---precursors of band gaps in larger samples. It is important to prove that the wave transport corresponds to strongly suppressed diffusion that is consistent with Anderson localization, in order to claim an Anderson mobility gap. This is achieved here by comparing experimental results with SC theory of localization.

To support our conclusions based on CBS measurements, we performed complementary experiments and analysis in transmission on sample L1. We used the technique of transverse confinement, which has been previously established as an unambiguous method of observing localization \cite{Hu2008}. The experimental method and comparison of measurements with SC theory have been presented in detail in Refs.\ \cite{Hu2008,PageVarenna2011,Hildebrand2015}. As can be seen in Fig.\ \ref{fig4}, the results  of transmission and reflection experiments agree reasonably well. From the combination of these measurements we estimate the position of MEs to be 1.198 $\pm$ 0.001 MHz and 1.243 $\pm$ 0.007 MHz. Inside the mobility gap the measured localization length reaches a minimum of 6.5 mm (3.8 times smaller than sample thickness).  The CBS results fluctuate much more with frequency, as do the CBS profiles themselves, especially around the upper ME where the position of the ME is less clear than for the lower ME.  While large fluctuations are to be expected in this regime, the precision of future measurements could be improved with a greater amount of configurational averaging, longer measurement times, and a wider angular array aperture.

Figure \ref{fig4} may be used to estimate the critical exponent of the localization transition $\nu$ because one expects $\xi(f) \propto |f - f_c|^{-\nu}$ for $f$ in the vicinity of a ME $f_c$. As can be seen in Fig.\ \ref{fig4}, $L/\xi$ looks approximately linear as a function of $f$ when it crosses the axis $L/\xi = 0$, leading to $\nu \approx 1$. It should be understood, however, that this result has large uncertainties due to the spread of data points in Fig.\ \ref{fig4} (especially at the upper ME). In addition, Fig.\ \ref{fig4} is obtained by fitting the experimental data with SC theory which is known to yield $\nu = 1$ in contradiction with numerical calculations \cite{Slevin2014} and may thus bias the result. More work is needed to obtain accurate estimates of $\nu$ for the localization transitions reported here.

In conclusion, we have employed the dynamic CBS effect to demonstrate an Anderson mobility gap in the spectrum of ultrasound scattered in a 3D strongly disordered elastic network. Performing our measurements in reflection instead of transmission as in previous works \cite{Hu2008,Hildebrand2014} ensured a sufficiently strong signal throughout the mobility gap, even for a very thick sample.  This is a significant advance, as previous experiments were only  able to reveal a single mobility edge \cite{Hildebrand2014}. Fits to the data by the self-consistent theory of localization yielded precisely the locations of the two mobility edges that serve as bounds of the mobility gap, and the localization length $\xi$ as a function of frequency. We were able to corroborate these results via transmission measurements on one of our samples. This work demonstrates the potential of dynamic CBS experiments to study localization effects in thick samples where transmission measurements are difficult or impossible, allowing us to access the deeply localized regime where $\xi \ll L$.  The thickness-independence of backscattering in a wide range of times provides an important advantage in the investigation of critical behavior where the elimination of finite-size effects is desired. This approach, made possible by a combination of modern experimental techniques with a careful theoretical description, can be extended to other classical waves (light, microwaves) as well.

\begin{acknowledgments}
We thank the Agence Nationale de la Recherche for financial support under grant ANR-14-CE26-0032 LOVE and CNRS for support in the framework of a France-Canada PICS project Ultra-ALT. J.H.P. and L.A.C. acknowledge the support of NSERC (Discovery Grant RGPIN/9037-2001, Canada Government Scholarship, and Michael Smith Foreign Study Supplement), the Canada Foundation for Innovation and the Manitoba Research and Innovation Fund (CFI/MRIF, LOF Project 23523). A.A. and A.D. benefited from funding by LABEX WIFI (Laboratory of Excellence ANR-10-LABX-24), within the French Program �Investments for the Future� under Reference No. ANR-10-IDEX-0001-02 PSL*.
\end{acknowledgments}

\newpage

\renewcommand{\thefigure}{S\arabic{figure}}
\renewcommand{\theequation}{S\arabic{equation}}

\setcounter{equation}{0}
\setcounter{figure}{0}
\bibliographystyle{apsrev4-1}
\renewcommand*{\citenumfont}[1]{S#1}
\renewcommand*{\bibnumfmt}[1]{[S#1]}

\begin{center}
\Large{\bf{Supplemental material}}
\end{center}
\normalsize

\section{Introduction}
This document provides further information on the experimental techniques, the self-consistent theory for backscattered intensity, and the procedure used to fit the predictions of this model to experimental data.  Representative values of the best-fit parameters are also presented and discussed.

\section{Experimental method} 

In this section, we give additional details on the backscattering experiments that we performed to demonstrate a robust new approach for investigating 3D Anderson localization. As emphasized in the letter, time- and angle-resolved backscattering experiments have several important advantages compared with the transmission measurements used in previous studies, enabling investigations of Anderson localization all the way through any mobility gap. Access to the deeply localized regime, where $\xi<<L$, obviously requires that the signals emerging from the medium be large enough to be measurable. In transmission, this requirement is difficult, if not impossible, to satisfy. In previous works, transmission through the samples was so greatly reduced inside the transmission dips (where a mobility edge was demonstrated) that measurements were not possible all of the way through the mobility gap, the most deeply localized regime was inaccessible, and the upper mobility edge could not be identified \cite{Hildebrand2014SM,Hildebrand2015SM}. By contrast, the reflection geometry that we employ here capitalizes on the distinct advantage that backscattered ultrasound is not affected by this limitation, allowing arbitrarily thick samples to be studied, and a complete investigation of the entire localization regime to be carried out. In addition, backscattering measurements are independent of sample thickness over a significant range of times before the detected signals have been able to reach and travel back from the far side of the sample.  This not only simplifies the interpretation of the current backscattering measurements but will also enable future investigations of critical behaviour in which finite size effects can be eliminated.  It is these considerations that motivated the design of our 
backscattering experiments, and have led to the significant progress in the investigation of 3D Anderson localization that is highlighted in the conclusions of our letter.

Given these advantages of backscattering measurements, one might wonder why we have focused on dynamic coherent backscattering rather than near-field detection of the time-dependent transverse intensity profile at the surface of the sample. Such dynamic transverse profile measurements would be expected to give the same type of (absorption-free) information on localization as was obtained previously in transmission \cite{Hu2008SM}, but with all of the additional advantages of the reflection geometry. While this is true in principle, we found that practical limitations preclude effective measurements of this type in reflection. Specifically, near-field measurements in reflection are extremely problematic because the placement of transducers at the sample surface leads to spurious reflections between the generator, sample surface, and detector. In addition, the generation and detection transducers get in the way of each other, making measurements difficult and data for some positions simply inaccessible. We also tried making measurements of the near-field transverse profile through the use of ultrasonic arrays in direct contact with the sample surface, but these were plagued by crosstalk between transducer elements during emission, which interfered with the detection of the interesting signals that have penetrated inside the sample. In addition, placing an array in contact with the sample complicates the boundary conditions. In contrast, coherent backscattering enables the spatial Fourier transform of the entire spatial intensity profile to be measured \emph{in the far-field} with a single ultrasonic transducer array, making it the perfect tool to investigate the growth (or not) of the transverse width in reflection.

The backscattering experiments, as well as the transmission measurements used to corroborate the results for sample L1, were carried out by immersing water-proofed samples and transducers in a large water tank.  The pores between the brazed beads in the samples were held under vacuum, thus ensuring that ultrasonic transport inside the sample was confined to the elastic bead network, and that both backscattering and transmission experiments were performed under the same conditions (apart from placement and type of ultrasonic emitters and detectors used). Thus, although both longitudinal and transverse elastic waves are present inside our solid samples, the emitted and measured signals for all experiments have longitudinal polarization (acoustic waves in water).

In backscattering, the response matrix was measured for sample L1 (L2) using 64 (128) elements of a linear ultrasonic array with a central frequency of 1.6 (1.0) MHz, capable of emitting/detecting signals for a frequency range of 0.6 - 1.9 MHz (0.5 - 1.4 MHz). Utmost care was taken to ensure that all possible contributions due to stray background signals were eliminated from the backscattering data by systematically searching for such contributions, removing them where possible, and analysing the data only over the range of times where valid data, uncontaminated by stray signals, were detected.  For example, careful placement of the array and sample, the design of a support system for the sample that eliminated spurious reflections, as well as checks with (temporarily inserted) reflecting or opaque objects, were used to ensure that effects from the edges of samples were negligible. The use of short pulses and a large water tank ensured that reflections from the sides of the tank arrived after the backscattered signals from the sample. The data were analyzed only for times greater than 40 $\mu$s (to discard any vestiges of specular reflections from the sample surface and single scattering, which might have persisted despite the sophisticated filtering technique that were used to remove these contributions \cite{Aubry2014SM}) and for times less than 200 (120) $\mu$s for L1 (L2) (to reject contributions from echoes between the array and sample). Similar care was employed to ensure that only multiply scattered signals from inside the sample were analysed for the transmission measurements on L1 (see Refs. \cite{Hildebrand2014SM,Hildebrand2015SM} for details on similar transmission experiments).

For configurational averaging of the backscattering data, the array was translated parallel to the sample, acquiring response matrices at 302 (66) different positions. The distance between the array and sample was 182 (136) mm, so that the backscattering experiments were carried in out in the far field, which for diffuse waves is defined by the condition $a \gg \sqrt{D_B t}$ ($a$ is the sample-array distance, $D_B$ is the Boltzmann diffusion coefficient, $t$ is time). In the diffuse regime (e.g.{,} 1.65 MHz for sample L1), $D_B$ for sample L1 has been measured to be approximately 0.7 $\text{mm}^2/\mu $s, and the longest times experimentally available to us are 210 $\mu$s, so the approximation of $a = 182 \text{ mm } \gg \sqrt{0.7 \times 210} \approx 12$ mm is valid. In the localized regime, the dynamic spreading of the diffuse halo is less, so that the far-field limit is even better respected.

\begin{figure*}[t]
\centering
\includegraphics[width=\textwidth]{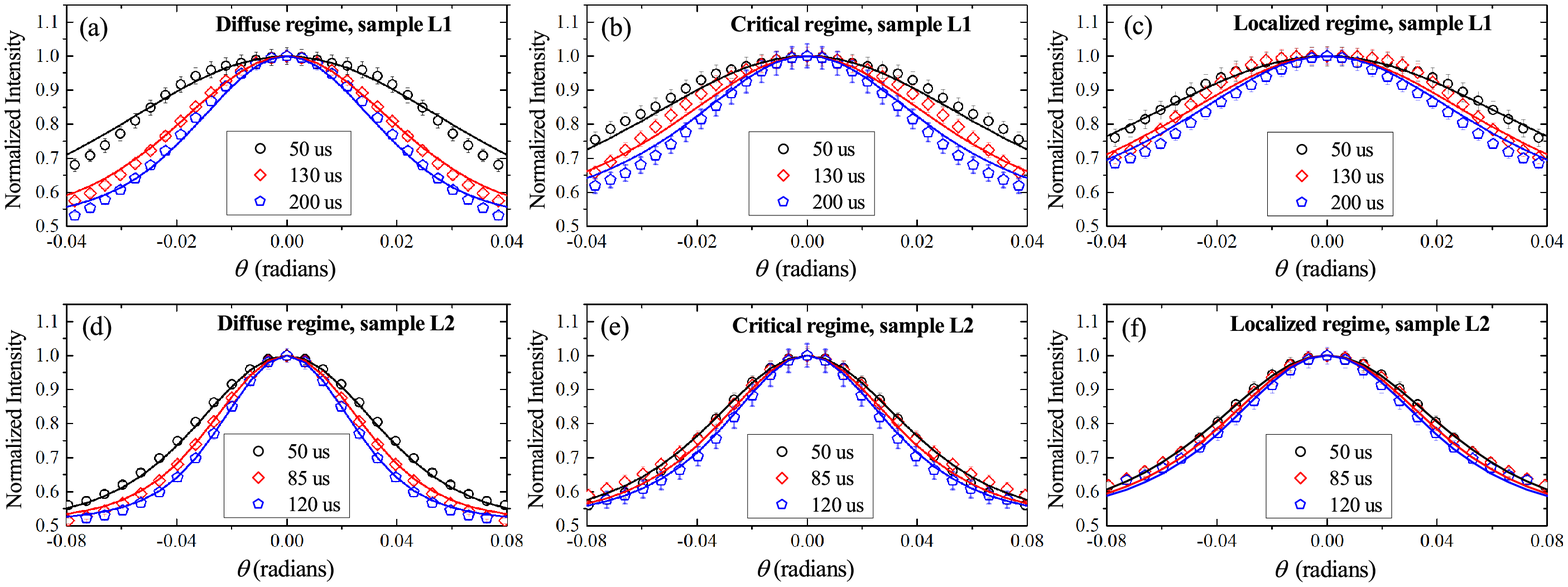}
\caption{Dynamic CBS profiles from samples L1 (top) and L2 (bottom) for three different times, and for three different frequencies:
(a),(d) 1.18 MHz in the diffuse regime (correlation length $\xi = 2.1\pm0.2$ mm for L1 and $3.2\pm0.9$ mm for L2); (b),(e) near 1.20 MHz at a ME ($\xi$ diverges); and (c),(f) 1.22 MHz in the {localized} regime ($\xi = 12\pm1$ mm for L1 and $16\pm3$ mm for L1).  Solid lines are best fits of {SC theory} to the data (symbols). Note that the horizontal scales are different between (a,b,c) and (d,e,f), and that a different range of times is presented. It is also important to note that $\xi$ should not necessarily be the same for both samples at exactly the same frequency.}
\label{fig1S}
\end{figure*}

After filtering the recurrent scattering contribution, the bandwidth-limited time-dependent CBS profiles $R(\theta,t)$ were extracted from the conventional multiple scattering contribution to the response matrix.  The dynamic CBS profiles were normalized to eliminate the influence of absorption by dividing $R(\theta,t)$ by $R(0,t)$, since at time $t$ the effect of absorption on the numerator and denominator of this ratio is the same, and therefore cancels.  Typical results near the lower mobility edge are shown in Fig.~\ref{fig1S}, where the data are compared with theoretical predictions as described in the next two sections.

\section{Self-consistent theory for dynamic coherent backscattering}

Our theoretical model to describe the dynamic coherent backscattering (CBS) of ultrasound is based on the equations of self-consistent (SC) theory of Anderson localization with a position- and frequency-dependent diffusion coefficient $D(\mathbf{r}, \Omega)$ as derived in Ref.\ \cite{Cherroret2008SM}. In these equations, the scattering mean free path $\ell$ should be replaced by $\ell^*_B$---the transport mean free path in the absence of localization effects--- to account for the scattering anisotropy of our samples ($\ell^*_B > \ell$).

To define the mobility edge (ME) and the localization length, we first analyze SC equations in the infinite 3D medium where $D$ becomes independent of position. For the stationary ($\Omega = 0$) diffusion coefficient we obtain
\begin{eqnarray}
D = D_B \left[1 - \frac{3 \mu}{(k \ell^*_B)^2} \right],
\label{d0perp}
\end{eqnarray}
where $D_B$ is the (Boltzmann) diffusion coefficient in the absence of localization effects and an upper cut-off $q_{\perp}^{\mathrm{max}} = \mu/\ell^*_B$ (with $\mu \sim 1$) was introduced in the integration over the transverse momentum $q_{\perp} = \{ q_x, q_y \}$ in order to regularize the integral. Here we break the symmetry between $q_{\perp}$ and $q_z$ to anticipate the experimental geometry of a disordered slab perpendicular to the $z$ axis. A ME of the Anderson transition at $k \ell = (k\ell)_c$ corresponds to $\mu = \frac13 (k\ell)_c^2 (\ell^*_B/\ell)^2$. In the localized regime $k \ell < (k \ell)_c$, an analytic solution of the equations of SC theory can be obtained for a point source emitting a short pulse at $\mathbf{r}' = 0$ and $t' = 0$, in the long-time limit. We obtain an intensity Green's function
\begin{eqnarray}
C(\mathbf{r}, \mathbf{r}', t \to \infty) = \frac{1}{4 \pi \xi^2 |\mathbf{r}-\mathbf{r}'|}
\exp\left(-|\mathbf{r}-\mathbf{r}'|/\xi \right),\;\;\;\;
\label{longtime}
\end{eqnarray}
where the localization length is
\begin{eqnarray}
\xi = \frac{6 \ell}{(k \ell)_c^2} \left( \frac{\ell}{\ell^*_B} \right)
\frac{p^2}{1 - p^4},
\label{xi}
\end{eqnarray}
and $p = k\ell/(k\ell)_c$.

To describe the experimental data, we solve the equations of SC theory in a slab of thickness $L$ with boundary conditions derived in Ref.\ \cite{Cherroret2008SM}, where the extrapolation length
\begin{eqnarray}
z_0 = \frac23 \ell^*_B \frac{1+R}{1-R}
\label{z0}
\end{eqnarray}
depends on the internal reflection coefficient $R$.
To this end, we Fourier transform the SC equations in the transverse plane $\boldsymbol{\rho} = \{ x, y \}$ and discretize the remaining ordinary differential equation for $C(q_{\perp}, z, z', \Omega)$ on a grid for $z \in [0, L]$ \footnote{The actual calculation is performed in dimensionless variables, but we omit this technical detail here for the purpose of clarity.}. A sufficiently fine discretization is also introduced for $q_{\perp}$ and $\Omega$, and the resulting system of linear equations with a tridiagonal matrix of coefficients is solved numerically using a standard routine \verb?zgtsl? from LAPACK library \cite{lapack} for $D(z, \Omega) = D_B$. A new value of $D(z, \Omega)$ is then obtained from the return probability
\begin{eqnarray}
C(\mathbf{r}, \mathbf{r}' = \mathbf{r}, \Omega) = \frac{1}{2 \pi} \int\limits_0^{q_{\perp}^{\mathrm{max}}} d q_{\perp} q_{\perp} C(q_{\perp}, z, z'=z, \Omega),\;\;\;\;\;\;
\label{return}
\end{eqnarray}
and the solution is iterated until convergence, i.e., $D(z,\Omega)$ does not change by more than a very small amount, typically less than ($5$x$10^{-5}$)\%, from one iteration to the next. Transmission and reflection coefficients $T(q_{\perp}, \Omega)$ and $R(q_{\perp}, \Omega)$ are then calculated as
\begin{eqnarray}
R(q_{\perp}, \Omega) &=& D(z, \Omega)
\left. \frac{\partial}{\partial z} C(q_{\perp}, z, z'=\ell^*_B, \Omega) \right|_{z=0}\;\;\;
\label{refl}
\end{eqnarray}
and similarly for $T(q_{\perp}, \Omega)$. We obtain the time-dependent intensity profiles in transmission $T(\boldsymbol{\rho}, t)$ and reflection $R(\boldsymbol{\rho}, t)$ by a double inverse Fourier transform of $T(q_{\perp}, \Omega)$ and $R(q_{\perp}, \Omega)$, respectively. The dynamic CBS profile $R(\theta, t) = R(q_{\perp} = k_0 \sin\theta, t)$ follows from the observation that the CBS shape is given by the Fourier transform of the `diffuse intensity halo' at the sample surface \cite{Akkermans2007SM}.

\section{Fitting self-consistent theory to experimental backscattering data}

The theory for $R(\theta,t)$ developed in the previous section is valid in the far field, where {$\sin\theta = q_{\perp}/k_0$. Here  $k_0 = 2 \pi f / v_0$ and $v_0 \approx 1500$ m/s is the speed of sound in water.} This is the appropriate limit for comparing with the experimental data, since our backscattering experiments are performed in the far-field (discussed in the first section of this Supplemental Material).

Near the localized regime, backscattered waves may spend a long time in a thick sample without reaching the far side. This means that for a range of times less than twice the typical time for waves to cross the sample, which can be estimated from the peak in the time-dependent transmission, the CBS effect is not sensitive to sample thickness. This is significant because calculations for very thick samples can be prohibitively time-consuming, so the modeling of backscattering in the localized regime is more convenient when there is no explicit dependence on sample thickness. In other words, theory for backscattered waves from a thin sample may also be used for a thicker sample, provided that the range of times investigated is short enough. Here we calculate SC theory for sample L1 and can compare it to experimental CBS profiles of both L1 and L2.

Most input parameters for the calculation of SC theory were determined from measurements performed in separate experiments and could thus be fixed in the fitting procedure. These (fixed) parameters are: scattering mean free path $\ell = 0.9$ mm, reflection coefficient $R = 0.67$, and wave vector $k = 2 \pi f / v_p$, with phase velocity $v_p = 2.8$ mm/$\mu$s, giving $k\ell = 2.7$ for $f=1.2$ MHz. The remaining parameter, the transport mean free path {$l^*_B = 4$ mm}, was determined from SC fitting of transverse confinement (transmission) data from sample L1.

The most important parameter involved in SC theory calculations of $R(\theta,t)$ in the vicinity of an Anderson transition is the localization (correlation) length $\xi$. As this parameter is unknown a priori, theoretical predictions for $R(\theta,t)$ are calculated for a  large range of $\xi$ values from the diffuse to localized regimes (and back again). These values of $\xi$ are determined {from $k\ell$ and its critical value at the transition $(k\ell)_c$ using Eq.\ (\ref{xi})}, with $k\ell$ fixed at the experimentally estimated value for $f = 1.2$ MHz. For each frequency $f$ of experimental data, the experimental CBS matrix $R(\theta,t)$ is fitted with every theory set. All fits are least-squares comparisons between the 2D matrices from experiment and theory, $R(\theta,t)$, using the reduced $\chi^2$ to determine the best-fit values of $\xi$. All times and $\theta$ values are fit simultaneously. This fitting procedure was performed with Wavemetrics software IGOR Pro. By finding the best-fit theory set for each $f$, the frequency-dependence of the localization (correlation) length $\xi(f)$ was determined.  This, in turn, enabled the locations of the two MEs, $f_{c1}$ and $f_{c2}$, to be determined (these are the frequencies where $\xi$ diverges).

Representative fit results for both samples are shown in Fig.~\ref{fig1S}, showing the quality of the fits in the diffuse regime at a frequency below the first localization transition [Fig.~\ref{fig1S}(a),(d)], at the first mobility edge [Fig. \ref{fig1S}(b),(e)], and in the mobility gap [Fig. \ref{fig1S}(c),(f)].  In all cases, the experimental data are well-described by the  SC theory: the narrowing of CBS profiles with time is reduced as the ME is reached [Fig. \ref{fig1S}(b),(e)], and in the {localized} regime CBS profiles change even less with time [Fig. \ref{fig1S}(c),(f)], with the width approaching a constant at long times.

The Boltzmann diffusion coefficient $D_B$ was a free fit parameter, yielding $D_B(f)$ after the entire fitting process. For sample L1, $D_B \approx 10 \pm 7$ mm$^2\mu$s$^{-1}$ below 1.24 MHz, and $D_B \approx 5 \pm 2$ mm$^2\mu$s$^{-1}$ above 1.24 MHz (from transmission and reflection measurements). The frequency-dependence of  $D_B(f)$ is supported by visual inspection of the time-dependence of the transmitted intensity in these regimes, and these values of $D_B$ are similar to the results of previous measurements in similar samples \cite{Hu2008SM,Hildebrand2015SM}. For sample L2, the fitting results gave values of $D_B$ ranging from $D_B \sim 13 - 61$ mm$^2\mu$s$^{-1}$ below 1.24 MHz (peaking in the {localized} regime) and $D_B \sim 9$ mm$^2\mu$s$^{-1}$ above 1.24 MHz. However, for such a thick sample as L2, the backscattering data are not very sensitive to $D_B$ over the experimentally accessible range of times, so that these estimates for sample L2 are not likely to be very accurate, although they are still consistent with surprisingly large values of $D_B$, as has been found for other samples in the {localized} regime.  Such values imply anomalously large values of the energy velocity $v_E$ \cite{Hu2008SM}, motivating future work to seek a theoretical understanding of $v_E$ in the {localized} regime.

The only other fit parameter was the background intensity level, which was also allowed to vary freely.  For both samples L1 and L2, and for almost all frequencies, best fits gave a background of within $10\%$ of the value of 0.5 which would be expected after the removal of the recurrent scattering contribution.  This variation of the background intensity results from the challenges of completely removing the recurrent scattering contribution, especially at early times where recurrent scattering dominates the backscattered intensity; by allowing the background intensity to be a free fit parameter, we were able to ensure that these background fluctuations did not  degrade the reliabilty of our determination of the frequency dependence of the localization (correlation) length.

\end{document}